\documentclass[pra,twocolumn,amsmath,amssymb]{revtex4}

\usepackage{graphicx}

\newcommand{\beq}{\begin{equation}}
\newcommand{\eeq}{\end{equation}}
\newcommand{\beqa}{\begin{eqnarray}}
\newcommand{\eeqa}{\end{eqnarray}}
\newcommand{\ket} [1] {\vert #1 \rangle}
\newcommand{\bra} [1] {\langle #1 \vert}

\newcommand{\proj}[1]{\ket{#1}\bra{#1}}

\newcommand{\opnorm}[1]{|\!|\!|#1|\!|\!|_2}

\newtheorem{lemma}{Lemma}

\begin{document}

\title{Quantum circuit implementation of the Hamiltonian versions of Grover's algorithm}
\author{J\'er\'emie Roland}
\author{Nicolas J. Cerf}
\affiliation{Ecole Polytechnique, CP 165/59,
Universit\'e Libre de Bruxelles, 1050 Brussels, Belgium}

\date{\today}

\begin{abstract}
We analyze three different quantum search algorithms, the traditional Grover's algorithm,
its continuous-time analogue by Hamiltonian evolution, and finally the quantum search by local adiabatic evolution.
We show that they are closely related algorithms in the sense that they all perform a rotation, at a constant angular
velocity, from a uniform superposition of all states to the solution state. This make it possible to implement the
last two algorithms by Hamiltonian evolution on a conventional quantum circuit, while keeping the quadratic speedup of Grover's original algorithm.
\end{abstract}

\maketitle

\section{Introduction}
While the standard paradigm of quantum computation uses quantum gates (i.~e., unitary operators) applied sequentially
on a quantum register, recent developments have introduced a new type of quantum algorithms where the state of
the quantum register evolves continuously in time under the action of some Hamiltonian. It includes, for instance,
the ``analog analogue'' of Grover's algorithm \cite{farh98} or the quantum
algorithms by adiabatic evolution that have been intensively studied lately \cite{farh00,vdam01}. It has been shown
that these Hamiltonian algorithms are genuinely quantum in the sense that they reproduce the quadratic speed-up of
Grover's algorithm (see, in particular, the {\em local} adiabatic version of Grover's algorithm \cite{vdam01,rola02}).
The purpose of this paper is, on one hand, to clarify the links between these Hamiltonian algorithms and their
conventional discrete equivalents, and, on the other hand, to show how they can be implemented on a traditional
quantum circuit. This second issue is important because it was never shown before how it could be done keeping the
quadratic speed-up of Grover's algorithm. It appears that all these algorithms take a very similar form in the high
dimension limit, which is particularly surprising for the case of the adiabatic search algorithm. Specifically, we see that
the mixing parameter (which measures the mixing between the initial and final Hamiltonians in the adiabatic search
algorithm) has to evolve in such a way that the instantaneous ground state rotates at a constant rate from the
initial to the final ground state. This makes the link fully explicit with Grover's original algorithm.

\section{Traditional Grover's algorithm}
First of all, let us briefly recall the principle of Grover's algorithm \cite{grov96,grov97}.
It is designed to solve the problem of finding
the values $x$ for which a function $f(x)$, usually called the ``oracle'', is equal to $1$
(while it vanishes everywhere else). As quantum gates have to be reversible, the quantum oracle must take the form:
\beq
O_f:\mathcal{H}_N\otimes\mathcal{H}_2\rightarrow\mathcal{H}_N\otimes\mathcal{H}_2:
\ket{x}\otimes\ket{y}\rightarrow\ket{x}\otimes\ket{y\oplus f(x)}
\eeq
where the $N$ candidate solutions $\ket{x}$ are taken as the basis states of the Hilbert space $\mathcal{H}_N$,
while $\oplus$ stands for the addition modulo 2. By considering the second register $\mathcal{H}_2$ as an
ancilla and preparing it in the state $\frac{1}{\sqrt{2}}[\ket{0}-\ket{1}]$,
the application of $O_f$ on both registers will be equivalent to the following unitary operation on the first one:
\beq
U_f:\mathcal{H}_N\rightarrow\mathcal{H}_N:\ket{x}\rightarrow(-1)^{f(x)}\ket{x}
\eeq
To clarify the notations, we will throughout this article restrict ourselves to the case where there is only one solution $x=m$
(our results may easily be generalized to the case of $M$ solutions, roughly speaking by replacing $N$ by $N/M$ in all the
formulas below). In this case, $f(m)=1$ while $f(x)=0\ (\forall x\neq m)$ and $U_f$ may be rewritten
\beq
U_f=I-2\proj{m}.
\eeq

Initially, we have no idea of what the solution could be, so we prepare the system in a uniform superposition of all possible solutions:
\beq
\ket{s}=\frac{1}{\sqrt{N}}\sum_{x=0}^{N-1}\ket{x}.
\eeq
This state may easily be obtained by applying an Hadamard transform $H$ on the $n=\log_2{N}$ qubits realizing the
quantum register $\mathcal{H}_N$, initially prepared in state $\ket{0}$. The algorithm will also require the following operation:
\beq\label{U0}
U_0=H^{\otimes n}(I-2\proj{0})H^{\otimes n}=I-2\proj{s}.
\eeq

Let us define as a Grover iteration the operator $G=-U_0 U_f$. Throughout this article, we will always
assume that $N\gg 1$ for simplicity reasons but also because the link between the different versions of the algorithm will appear
much more clearly. In that limit, it may be shown that the Grover iteration becomes a
simple rotation of angle $\frac{2}{\sqrt{N}}$ in the subspace spanned by $\ket{s}$ and $\ket{m}$. More precisely,
successive applications of $G$ on the initial state $\ket{s}$ will progressively make this state rotate to the solution
state $\ket{m}$:
\beqa
\ket{\psi^\text{dis}_j}=G^j\ket{s}&\approx&\cos{\frac{2j}{\sqrt{N}}}\ket{s}+\sin{\frac{2j}{\sqrt{N}}}\ket{m}\\
&\approx&\cos\alpha_j^\text{dis}\ket{s}+\sin\alpha_j^\text{dis}\ket{m}
\eeqa
where
\beq
\alpha_j^\text{dis}=\frac{2j}{\sqrt{N}}.
\eeq
Thus, by applying the Grover iteration $R^\text{dis}\approx\frac{\pi}{4}\sqrt{N}$ times, we obtain the solution state $|m\rangle$ with a
probability close to $1$ with a quadratic speed-up with respect to a classical search, which would necessarily require a number of calls to the oracle $f(x)$ of order $N$.

Let us notice that as $\frac{\pi}{4}\sqrt{N}$ is generally not an integer, we have to round it, for instance to the nearest lower integer,
so that $R^\text{dis}=\lfloor\frac{\pi}{4}\sqrt{N}\rfloor$. This results in an error
\beq
\|\ket{\psi_{R^\text{dis}}^\text{dis}}-\ket{m}\|<\sin \frac{2}{\sqrt{N}}\approx\frac{2}{\sqrt{N}}
\eeq
that tends to zero for $N\to\infty$.

\section{Hamiltonian equivalent of the oracle}\label{hamoracle}

\begin{figure}[htb]
{\unitlength=1.000000pt
\begin{picture}(155.00,60.00)(0.00,0.00)
\put(125.00,10.00){\line(1,0){10.00}}
\put(125.00,30.00){\line(1,0){10.00}}
\put(125.00,35.00){\line(1,0){10.00}}
\put(125.00,40.00){\line(1,0){10.00}}
\put(125.00,45.00){\line(1,0){10.00}}
\put(125.00,50.00){\line(1,0){10.00}}
\put(25.00,10.00){\line(-1,0){10.00}}
\put(25.00,30.00){\line(-1,0){10.00}}
\put(25.00,35.00){\line(-1,0){10.00}}
\put(25.00,40.00){\line(-1,0){10.00}}
\put(25.00,45.00){\line(-1,0){10.00}}
\put(25.00,50.00){\line(-1,0){10.00}}
\put(45.00,35.00){\line(1,0){60.00}}
\put(105.00,45.00){\line(-1,0){60.00}}
\put(75.00,10.00){\makebox(0.00,0.00){$U_t$}}
\put(115.00,30.00){\makebox(0.00,0.00){$O_f$}}
\put(35.00,30.00){\makebox(0.00,0.00){$O_f$}}
\put(155.00,10.00){\makebox(0.00,0.00){$\ket{0}$}}
\put(155.00,40.00){\makebox(0.00,0.00){$e^{-iH_ft}\ket{x}$}}
\put(0.00,10.00){\makebox(0.00,0.00){$\ket{0}$}}
\put(0.00,40.00){\makebox(0.00,0.00){$\ket{x}$}}
\put(85.00,10.00){\line(1,0){20.00}}
\put(45.00,10.00){\line(1,0){20.00}}
\put(45.00,30.00){\line(1,0){60.00}}
\put(45.00,40.00){\line(1,0){60.00}}
\put(45.00,50.00){\line(1,0){60.00}}
\put(125.00,0.00){\line(-1,0){20.00}}
\put(125.00,60.00){\line(0,-1){60.00}}
\put(105.00,60.00){\line(1,0){20.00}}
\put(105.00,0.00){\line(0,1){60.00}}
\put(65.00,20.00){\line(0,-1){20.00}}
\put(85.00,20.00){\line(-1,0){20.00}}
\put(85.00,0.00){\line(0,1){20.00}}
\put(65.00,0.00){\line(1,0){20.00}}
\put(45.00,60.00){\line(-1,0){20.00}}
\put(45.00,0.00){\line(0,1){60.00}}
\put(25.00,0.00){\line(1,0){20.00}}
\put(25.00,60.00){\line(0,-1){60.00}}
\end{picture}}
\caption{Circuit for implementing the evolution of a Hamiltonian $H_f$ during a time $t$ by using twice the corresponding oracle $O_f$.\label{circuit}}
\end{figure}
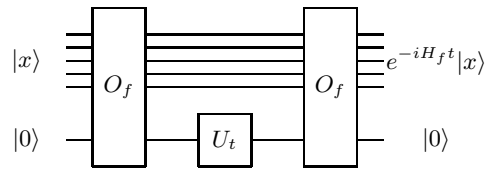

In the two Hamiltonian quantum search algorithms discussed below, we will use the Hamiltonian:
\beq
H_f=I-\proj{m}.
\eeq
Let us show why it can be considered as equivalent to the oracle $U_f$.
If we apply this Hamiltonian on a basis state $\ket{x}$ during a time $t$, it yields
\beqa
e^{-iH_ft}|x\rangle&=&
\left\{\begin{array}{ll}
|x\rangle & x=m\\
e^{-it}|x\rangle & \forall x\neq m.
\end{array}\right.\\
&=&e^{-i(1-f(x))t}|x\rangle.
\eeqa
We immediately see that by taking $t=\pi$, we reproduce the operation $-U_f$, that is
\beq\label{uf}
e^{-iH_f\pi}=-U_f
\eeq
Conversely, it is possible to simulate the application of $H_f$ during a time $t$ by a quantum circuit using a one-qubit
ancilla prepared in state $\ket{0}$, two calls to the oracle $O_f$ and an additional phase gate
\beq
U_t=e^{-it}\proj{0}+\proj{1}.
\eeq
Considering the circuit represented in Fig.~\ref{circuit}),
we have
\beqa
O_f\left[\ket{x}\otimes\ket{0}\right]&=&\ket{x}\otimes\ket{f(x)}\nonumber\\
I\otimes U_t\left[\ket{x}\otimes\ket{f(x)}\right]&=&e^{-i(1-f(x))t}\ket{x}\otimes\ket{f(x)}\nonumber\\
O_f\left[e^{-i(1-f(x))t}\ket{x}\otimes\ket{f(x)}\right]&=&e^{-i(1-f(x))t}\ket{x}\otimes\ket{0}\nonumber\\
&=&e^{-iH_ft}|x\rangle\otimes|0\rangle.
\eeqa

\section{Analog quantum search}
Let us now consider the ``analog'' algorithm introduced by Farhi et al. in \cite{farh98}. In addition to the
oracle Hamiltonian $H_f$, we will need a second Hamiltonian
\beq
H_0=H^{\otimes n}(I-\proj{0})H^{\otimes n}=I-\proj{s}
\eeq
that is related to $U_0$ as defined in Eq.~(\ref{U0}) in the same manner than $H_f$ is related to $U_f$. The algorithm consists in preparing the system
in the starting state $\ket{\psi^\text{an}(t=0)}=\ket{s}$ and then let it evolve under the time-independent Hamiltonian
$H^\text{an}=H_0+H_f$. We may show by simple calculation that:
\beqa
\ket{\psi^\text{an}(t)}&=&e^{-iH^\text{an}t}\ket{s}\\
&=&e^{-it}[\cos{\frac{t}{\sqrt{N}}}\ket{s}+i\sin{\frac{t}{\sqrt{N}}}\ket{m}]\\
&=&e^{-it}[\cos\alpha^\text{an}(t)\ket{s}+i\sin\alpha^\text{an}(t)\ket{m}],\label{i}
\eeqa
where
\beq
\alpha^\text{an}(t)=\frac{t}{\sqrt{N}}.
\eeq
As in the traditional algorithm, the search works via a rotation from $\ket{s}$ to $\ket{m}$. However, the rotation is continuous here, and follows
a different path because of the presence of $i$ in the second term of Eq.~(\ref{i}). The solution state is thus
obtained with probability one if we apply $H^\text{an}$ during a time $T^\text{an}=\frac{\pi}{2}\sqrt{N}$.
Let us also notice that
\beqa
\alpha^\text{an}(2j)=\alpha^\text{dis}_j
\eeqa
which shows that the application of $H^\text{an}$ during a time $T^\text{an}/R^\text{dis}=2$ corresponds roughly to
one Grover iteration.
\par

Suppose now we want to implement this analog algorithm on a quantum circuit. We showed in the previous section how
to reproduce the application of $H_f$ with a circuit, but this does not allow us to directly apply
$H^\text{an}=H_0+H_f$. In order to achieve this, we need to cut the evolution time $T^\text{an}$ into $R^\text{an}$ small intervals
$\Delta T=\frac{T^\text{an}}{R^\text{an}}=\frac{\pi}{2}\frac{\sqrt{N}}{R^\text{an}}$ such that we may approximate
\beq
U(\Delta T)=e^{-i (H_0+H_f)\Delta T}
\eeq
by
\beq
U_{\Delta T}'=e^{-iH_0\Delta T} e^{-iH_f\Delta T}.
\eeq
Using the Campbell-Baker-Hausdorff approximation, which states that $\opnorm{e^{A+B}-e^Ae^B}\in O(\opnorm{[A,B]})$
with $\opnorm{A}=\max_{\|\ket{x}\|=1}\|A\ket{x}\|$ denoting the operator norm of $A$, we have
\beq
\opnorm{U(\Delta T)-U_{\Delta T}'}\in O(\opnorm{[H_0,H_f]}\Delta T^2).
\eeq
As
\beq
[H_0,H_f]=\sqrt{\frac{1}{N}}\sqrt{1-\frac{1}{N}}\lesssim\frac{1}{\sqrt{N}},
\eeq
the error introduced at each step is of order
\beq
\opnorm{U(\Delta T)-U_{\Delta T}'}\in O(\frac{\sqrt{N}}{{R^\text{an}}^2})
\eeq
Since there are $R^\text{an}$ successive steps, the global error made by this discretized analog algorithm is
\beq
\opnorm{U(T)-(U_{\Delta T}')^{R^\text{an}}}\in O(\frac{\sqrt{N}}{R^\text{an}}).
\eeq
This simply results from the property that if the condition $\opnorm{U_j-U'_j}\ll 1$ is fulfilled $\forall j$, then
\beq\label{opnorm}
\opnorm{\prod_jU_j-\prod_jU'_j}\leq\sum_j\opnorm{U_j-U'_j}.
\eeq
We thus observe that for a number of steps $R^\text{an}=\lfloor\frac{\sqrt{N}}{\epsilon}\rfloor$ (of the same order in $N$
as in Grover's traditional algorithm), we get the desired state with an error of order $\epsilon$.

Furthermore, keeping the results of the previous section
in mind, each step will have the same form $e^{-iH_0\delta T}e^{-iH_f\delta T}$ as a Grover iteration $G=-U_0U_f$,
and therefore will require $2$ calls to the oracle $O_f$ for being implemented with a quantum circuit.

\section{Quantum search by local adiabatic evolution}

For this third algorithm, exposed in \cite{vdam01} and \cite{rola02}, we will once more need the Hamiltonians
$H_0$ and $H_f$, and we will initially prepare our system in a uniform superposition of all possible solutions
$\ket{\psi^\text{ad}(t=0)}=\ket{s}$. This time, however, we apply a time-dependent Hamiltonian
\beq
H(s)=(1-s)H_0+s H_f
\eeq
to the system, where the mixing parameter $s=s(t)$ is a monotonic function with $s(0)=0$ and $s(T^\text{ad})=1$. As $\ket{s}$ is the ground state of $H(0)=H_0$,
the Adiabatic Theorem \cite{schiff} tells us that during the evolution $s(t)$, the system will stay near the instantaneous ground state
of $H(s)$ as long as the evolution of $H(s)$ is slow enough. If this condition is satisfied, the system will thus
end up in the ground state of $H(1)=H_f$, which is the solution state $\ket{m}$.
\par 

Let us first study the path followed by $\ket{\psi^\text{ad}(t)}$ during the evolution. As $H(s)$ only acts on the subspace spanned
by $\ket{s}$ and $\ket{m}$ and as we start from $\ket{s}$, the path followed by $\ket{\psi^\text{ad}(t)}$ will remain in this subspace
so that the problem may be studied in this $2$-dimensional Hilbert space. By calculating the eigenstates of $H(s)$, we find
\beqa
\ket{E_0;s}&=&\frac{\sqrt{N}(E_1(s)-s)\ket{s}+s\ket{m}}{\sqrt{E_1(s)^2+(N-1)(E_1(s)-s)^2}}\\
\ket{E_1;s}&=&\frac{\sqrt{N}(E_0(s)-s)\ket{s}+s\ket{m}}{\sqrt{E_0(s)^2+(N-1)(E_0(s)-s)^2}}
\eeqa
where
\beqa
E_0(s)&=&\frac{1}{2}\left[1-\sqrt{1-4\frac{N-1}{N}s(1-s)}\right]\\
E_1(s)&=&\frac{1}{2}\left[1+\sqrt{1-4\frac{N-1}{N}s(1-s)}\right].
\eeqa
If the adiabatic condition is realized (see \cite{rola02}), we have
\beqa
\ket{\psi^\text{ad}(s)}&\approx&\frac{\sqrt{N}(E_1(s)-s)\ket{s}+s\ket{m}}{\sqrt{E_1(s)^2+(N-1)(E_1(s)-s)^2}}\\
&\approx&\cos\alpha^\text{ad}(s)\ket{s}+\sin\alpha^\text{ad}(s)\ket{m}
\eeqa
where
\beqa
\alpha^\text{ad}(s)&=&\arctan \frac{s}{\sqrt{N}(E_1(s)-s)}\\
&\approx&\frac{1}{2}\arctan\frac{2s}{\sqrt{N}(1-2s)}
\eeqa
in the limit $N\gg1$.
\begin{figure}[htb]
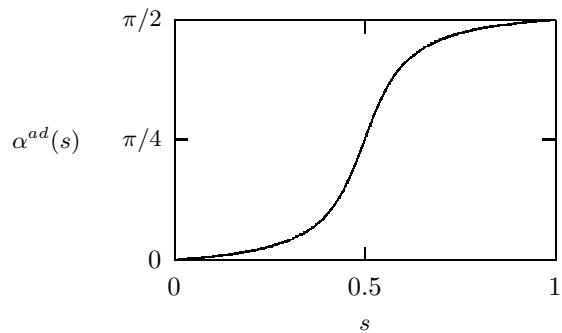

\input figure2.tex
\caption{Rotation angle $\alpha^\text{ad}(s)$ for the adiabatic quantum search algorithm with $N=32$.\label{graphalpha}}
\end{figure}

The function $\alpha^\text{ad}(s)$ is plotted in Fig.~\ref{graphalpha}. We see that the evolution is once again a rotation
from $\ket{s}$ to $\ket{m}$, but which is {\em not} performed at a constant angular velocity if $s(t)$ is chosen to be linear in $t$,
which corresponds to the quantum search by {\em global} adiabatic evolution originally described in \cite{farh00}.
The observed angular velocity is indeed greater for $s$ close to $1/2$ while it is smaller at the beginning and the end of the time evolution.
Let us also notice that at discrete values $s(t)=s_k$, the continuous path $\ket{\psi^\text{ad}(s)}$ coincides with the states $\ket{\psi^\text{dis}_j}$
of Grover's traditional algorithm. Thus, in the global adiabatic search  algorithm, the system exactly follows the path of Grover's algorithm,
but at a varying rate. This suggests that this algorithm is not the correct adiabatic equivalent to Grover's algorithm. Moreover, we note that
if $s(t)=t/T^\text{ad}$, then the Adiabatic Theorem imposes that $T^\text{ad}\in O(N)$, so that we loose the quadratic speed-up of Grover's algorithm (see \cite{rola02}).

In order to circumvent this problem, we can perform a {\em local} adiabatic evolution as defined in \cite{rola02}.
Then we get the solution state with an error less than $\epsilon$
\beq
\|\ket{\psi^\text{ad}(T^\text{ad})}-\ket{m}\|\leq\epsilon
\eeq
provided that we evolve at a rate such that
\beqa
t(s)&=&\frac{N}{2\epsilon\sqrt{N-1}}[\arctan [\sqrt{N-1} (2s-1)]\nonumber\\
&&+\arctan\sqrt{N-1}]\nonumber\\
&=&\frac{N}{2\epsilon\sqrt{N-1}}\arctan \frac{\sqrt{N-1}(2s-1)+\sqrt{N-1}}{1-(N-1)(2s-1)}\nonumber\\
&\approx&\frac{\sqrt{N}}{2\epsilon}\arctan\frac{2s}{\sqrt{N}(1-2s)}
\eeqa
in the limit $N\gg 1$. Thus, for a local adiabatic evolution,
\beq
\alpha^\text{ad}(t)\approx\frac{\epsilon t}{\sqrt{N}}.
\eeq
This is now a rotation at a constant rate during a time $T^\text{ad}=\frac{\pi}{2\epsilon}\sqrt{N}$, so that we may consider
this local adiabatic evolution as the right equivalent to Grover's algorithm.

Let us study the implementation of this algorithm on a quantum circuit (we will closely follow the lines of the
development exposed in \cite{vdam01}). As for the analog quantum search, we discretize
the evolution by cutting the time $T^\text{ad}$ in $R^\text{ad}$ intervals $\Delta T=\frac{T^\text{ad}}{R^\text{ad}}$.
During each interval $[t_{j-1},t_j] (t_j=j\Delta T)$, we approximate the varying Hamiltonian
$H(s(t))$ by the constant one $H_j=(1-s_j) H_0+s_j H_f\quad(s_j=s(t_j))$. In order to evaluate the error introduced by this
approximation, we will use the following lemma:
\begin{lemma}
Let $H(t)$ and $H'(t)$ be two time-dependent Hamiltonians for $0\leq t\leq T$, and let $U(T)$ and $U'(T)$ be
the respective unitary evolutions that they induce. If the difference between the Hamiltonians is limited by
$\opnorm{H(t)-H'(t)}\leq\delta(t)$, then the distance between the induced transformations is bounded by
$\opnorm{U(T)-U'(T)}\leq\sqrt{2\int_0^T\delta(t)dt}$.
\end{lemma}
This lemma is a straightforward generalization of the one introduced
in \cite{vdam01}, with the important difference that the Hamiltonian
difference $\delta(t)$ may vary in time. This will be crucial in order to keep the quadratic speedup of Grover's search after
this discretization procedure (this is reminiscent to the distinction between the global and the local adiabatic evolution).
The proof of Lemma 1 is left to the reader.

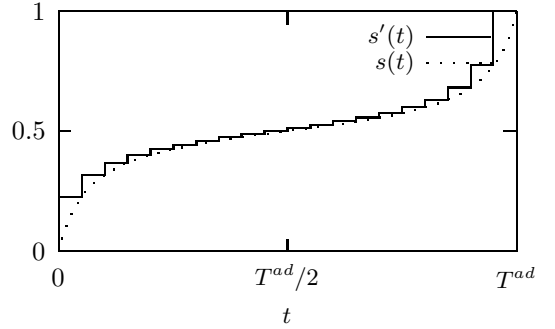
\begin{figure}[htb]
{\setlength{\unitlength}{0.240900pt}
\ifx\plotpoint\undefined\newsavebox{\plotpoint}\fi
\sbox{\plotpoint}{\rule[-0.200pt]{0.400pt}{0.400pt}}%
\begin{picture}(900,540)(0,0)
\font\gnuplot=cmr10 at 10pt
\gnuplot
\sbox{\plotpoint}{\rule[-0.200pt]{0.400pt}{0.400pt}}%
\put(120.0,123.0){\rule[-0.200pt]{4.818pt}{0.400pt}}
\put(100,123){\makebox(0,0)[r]{0}}
\put(819.0,123.0){\rule[-0.200pt]{4.818pt}{0.400pt}}
\put(120.0,312.0){\rule[-0.200pt]{4.818pt}{0.400pt}}
\put(100,312){\makebox(0,0)[r]{0.5}}
\put(819.0,312.0){\rule[-0.200pt]{4.818pt}{0.400pt}}
\put(120.0,500.0){\rule[-0.200pt]{4.818pt}{0.400pt}}
\put(100,500){\makebox(0,0)[r]{1}}
\put(819.0,500.0){\rule[-0.200pt]{4.818pt}{0.400pt}}
\put(120.0,123.0){\rule[-0.200pt]{0.400pt}{4.818pt}}
\put(120,82){\makebox(0,0){0}}
\put(120.0,480.0){\rule[-0.200pt]{0.400pt}{4.818pt}}
\put(479.0,123.0){\rule[-0.200pt]{0.400pt}{4.818pt}}
\put(479,82){\makebox(0,0){$T^{ad}/2$}}
\put(479.0,480.0){\rule[-0.200pt]{0.400pt}{4.818pt}}
\put(839.0,123.0){\rule[-0.200pt]{0.400pt}{4.818pt}}
\put(839,82){\makebox(0,0){$T^{ad}$}}
\put(839.0,480.0){\rule[-0.200pt]{0.400pt}{4.818pt}}
\put(120.0,123.0){\rule[-0.200pt]{173.207pt}{0.400pt}}
\put(839.0,123.0){\rule[-0.200pt]{0.400pt}{90.819pt}}
\put(120.0,500.0){\rule[-0.200pt]{173.207pt}{0.400pt}}
\put(479,21){\makebox(0,0){$t$}}
\put(120.0,123.0){\rule[-0.200pt]{0.400pt}{90.819pt}}
\put(679,460){\makebox(0,0)[r]{$s'(t)$}}
\put(699.0,460.0){\rule[-0.200pt]{24.090pt}{0.400pt}}
\put(120,123){\usebox{\plotpoint}}
\put(120.0,123.0){\rule[-0.200pt]{0.400pt}{20.476pt}}
\put(120.0,208.0){\rule[-0.200pt]{8.672pt}{0.400pt}}
\put(156.0,208.0){\rule[-0.200pt]{0.400pt}{8.431pt}}
\put(156.0,243.0){\rule[-0.200pt]{8.672pt}{0.400pt}}
\put(192.0,243.0){\rule[-0.200pt]{0.400pt}{4.577pt}}
\put(192.0,262.0){\rule[-0.200pt]{8.672pt}{0.400pt}}
\put(228.0,262.0){\rule[-0.200pt]{0.400pt}{2.891pt}}
\put(228.0,274.0){\rule[-0.200pt]{8.672pt}{0.400pt}}
\put(264.0,274.0){\rule[-0.200pt]{0.400pt}{2.168pt}}
\put(264.0,283.0){\rule[-0.200pt]{8.672pt}{0.400pt}}
\put(300.0,283.0){\rule[-0.200pt]{0.400pt}{1.686pt}}
\put(300.0,290.0){\rule[-0.200pt]{8.672pt}{0.400pt}}
\put(336.0,290.0){\rule[-0.200pt]{0.400pt}{1.445pt}}
\put(336.0,296.0){\rule[-0.200pt]{8.431pt}{0.400pt}}
\put(371.0,296.0){\rule[-0.200pt]{0.400pt}{1.445pt}}
\put(371.0,302.0){\rule[-0.200pt]{8.672pt}{0.400pt}}
\put(407.0,302.0){\rule[-0.200pt]{0.400pt}{1.204pt}}
\put(407.0,307.0){\rule[-0.200pt]{8.672pt}{0.400pt}}
\put(443.0,307.0){\rule[-0.200pt]{0.400pt}{1.204pt}}
\put(443.0,312.0){\rule[-0.200pt]{8.672pt}{0.400pt}}
\put(479.0,312.0){\rule[-0.200pt]{0.400pt}{0.964pt}}
\put(479.0,316.0){\rule[-0.200pt]{8.672pt}{0.400pt}}
\put(515.0,316.0){\rule[-0.200pt]{0.400pt}{1.204pt}}
\put(515.0,321.0){\rule[-0.200pt]{8.672pt}{0.400pt}}
\put(551.0,321.0){\rule[-0.200pt]{0.400pt}{1.445pt}}
\put(551.0,327.0){\rule[-0.200pt]{8.672pt}{0.400pt}}
\put(587.0,327.0){\rule[-0.200pt]{0.400pt}{1.445pt}}
\put(587.0,333.0){\rule[-0.200pt]{8.672pt}{0.400pt}}
\put(623.0,333.0){\rule[-0.200pt]{0.400pt}{1.686pt}}
\put(623.0,340.0){\rule[-0.200pt]{8.672pt}{0.400pt}}
\put(659.0,340.0){\rule[-0.200pt]{0.400pt}{2.168pt}}
\put(659.0,349.0){\rule[-0.200pt]{8.672pt}{0.400pt}}
\put(695.0,349.0){\rule[-0.200pt]{0.400pt}{2.891pt}}
\put(695.0,361.0){\rule[-0.200pt]{8.672pt}{0.400pt}}
\put(731.0,361.0){\rule[-0.200pt]{0.400pt}{4.577pt}}
\put(731.0,380.0){\rule[-0.200pt]{8.672pt}{0.400pt}}
\put(767.0,380.0){\rule[-0.200pt]{0.400pt}{8.431pt}}
\put(767.0,415.0){\rule[-0.200pt]{8.431pt}{0.400pt}}
\put(802.0,415.0){\rule[-0.200pt]{0.400pt}{20.476pt}}
\put(802.0,500.0){\rule[-0.200pt]{8.672pt}{0.400pt}}
\put(679,419){\makebox(0,0)[r]{$s(t)$}}
\multiput(699,419)(20.756,0.000){5}{\usebox{\plotpoint}}
\put(799,419){\usebox{\plotpoint}}
\put(120,123){\usebox{\plotpoint}}
\put(120.00,123.00){\usebox{\plotpoint}}
\put(125.01,143.02){\usebox{\plotpoint}}
\put(132.16,162.47){\usebox{\plotpoint}}
\put(140.19,181.39){\usebox{\plotpoint}}
\put(150.08,199.08){\usebox{\plotpoint}}
\put(161.48,215.48){\usebox{\plotpoint}}
\put(175.74,229.74){\usebox{\plotpoint}}
\put(190.46,241.46){\usebox{\plotpoint}}
\put(206.48,251.48){\usebox{\plotpoint}}
\put(222.68,259.68){\usebox{\plotpoint}}
\put(240.16,266.16){\usebox{\plotpoint}}
\put(258.35,272.35){\usebox{\plotpoint}}
\put(276.42,277.42){\usebox{\plotpoint}}
\put(294.69,282.00){\usebox{\plotpoint}}
\put(313.21,286.00){\usebox{\plotpoint}}
\put(332.30,290.00){\usebox{\plotpoint}}
\put(351.23,293.00){\usebox{\plotpoint}}
\put(370.16,296.00){\usebox{\plotpoint}}
\put(389.67,299.00){\usebox{\plotpoint}}
\put(409.18,302.00){\usebox{\plotpoint}}
\put(428.70,305.00){\usebox{\plotpoint}}
\put(448.62,307.00){\usebox{\plotpoint}}
\put(468.14,310.00){\usebox{\plotpoint}}
\put(487.65,313.00){\usebox{\plotpoint}}
\put(507.58,315.00){\usebox{\plotpoint}}
\put(527.09,318.00){\usebox{\plotpoint}}
\put(546.72,320.72){\usebox{\plotpoint}}
\put(566.53,323.00){\usebox{\plotpoint}}
\put(586.04,326.00){\usebox{\plotpoint}}
\put(604.97,329.00){\usebox{\plotpoint}}
\put(623.48,333.00){\usebox{\plotpoint}}
\put(643.00,336.00){\usebox{\plotpoint}}
\put(660.92,340.00){\usebox{\plotpoint}}
\put(679.00,344.44){\usebox{\plotpoint}}
\put(697.00,349.53){\usebox{\plotpoint}}
\put(715.15,355.15){\usebox{\plotpoint}}
\put(733.05,362.05){\usebox{\plotpoint}}
\put(750.36,370.36){\usebox{\plotpoint}}
\put(766.54,380.00){\usebox{\plotpoint}}
\put(781.82,391.82){\usebox{\plotpoint}}
\put(795.67,405.67){\usebox{\plotpoint}}
\put(807.93,421.86){\usebox{\plotpoint}}
\put(818.24,439.47){\usebox{\plotpoint}}
\put(826.58,458.16){\usebox{\plotpoint}}
\put(832.90,477.71){\usebox{\plotpoint}}
\put(838.00,497.67){\usebox{\plotpoint}}
\put(838,498){\usebox{\plotpoint}}
\end{picture}}
\caption{$s(t)$ and its discrete approximation $s'(t)$ (using $R^\text{ad}=20$ steps) for the local adiabatic algorithm with $N=32$.}
\label{graph}
\end{figure}

Thus, the approximation we made is equivalent to replacing the actual Hamiltonian $H(t)=H(s(t))$ by
$H'(t)=H(s'(t))$ where $s'(t)$ is a new monotonic
function approaching $s(t)$ but varying at times $t_j$ only (see Fig.~\ref{graph}).

We have
\beqa
\opnorm{H(t)-H'(t)}&=&\opnorm{H(s(t))-H(s'(t))}\\
&=&|s'(t)-s(t)|\ \opnorm{H_{in}-H_f}\\
&\leq& (\frac{ds}{dt}\Delta T +O(\Delta T^2) )\opnorm{H_{in}-H_f}.\nonumber
\eeqa
We may use Lemma 1 with $\delta(t)=\frac{ds}{dt}\Delta T \opnorm{H_{in}-H_f}$:
\beqa
\int_0^T \delta(t) dt&=&\Delta T \opnorm{H_{in}-H_f}\int_0^T \frac{ds}{dt} dt\\
&=&\Delta T \opnorm{H_{in}-H_f} \int_0^1 ds\\
&=&\Delta T \opnorm{H_{in}-H_f}\\
&\leq&\Delta T
\eeqa
as we may easily see that $\opnorm{H_{in}-H_f}\leq 1$. Now, the lemma gives:
\beq
\opnorm{U(T)-U'(T)}\leq\sqrt{2\Delta T}=\sqrt{2\frac{T^\text{ad}}{R^\text{ad}}}
\eeq
so that in order to keep an error of constant order $\epsilon$ for growing $N$, we must choose a number of steps
proportionnal to $T^\text{ad}$, that is $R^\text{ad}=\lfloor\frac{\sqrt{N}}{\epsilon^3}\rfloor$.

\begin{table*}
\begin{tabular}{|l|c|c|c|}
\hline
& Grover & Analog & Adiabatic\\
\hline
\#steps $R$ & $\lfloor\frac{\pi}{4}\sqrt{N}\rfloor$ & $\lfloor\frac{\sqrt{N}}{\epsilon}\rfloor$ & $\lfloor\frac{\sqrt{N}}{\epsilon^3}\rfloor$\\
Step $j$ & $\delta t_{0,j}=\delta t_{f,j}=\pi$ & $\delta t_{0,j}=\delta t_{f,j}=\epsilon\frac{\pi}{2}$
   & $\left\{\begin{array}{l}\delta t_{0,j}=(1-s_j)\epsilon^2\frac{\pi}{2}\\
   \delta t_{f,j}=s_j\epsilon^2\frac{\pi}{2}\end{array}\right.$\\
State $\ket{\psi_j}$ & $\cos\alpha_j\ket{s}+\sin\alpha_j\ket{m}$ & $\cos\alpha_j\ket{s}+i\sin\alpha_j\ket{m}$
   & $\cos\alpha_j\ket{s}+\sin\alpha_j\ket{m}$\\
Angle $\alpha_j$ & $\frac{2j}{\sqrt{N}}$ & $\frac{\epsilon j}{\sqrt{N}}\frac{\pi}{2}$ & $\frac{\epsilon^3 j}{\sqrt{N}}\frac{\pi}{2}$\\
Error $\|\ket{\psi_R}-\ket{m}\|$ & $O(\frac{1}{\sqrt{N}})$ & $O(\epsilon)$ & $O(\epsilon)$\\
\hline
\end{tabular}
\caption{Summary of the properties of the three quantum search algorithms. Here $\delta t_{0,j}$ and $\delta t_{f,j}$ are the times during which
$H_0$ and $H_f$ have to be applied in step $j$. Notice that we have omitted some irrelevant global phases in front of $\ket{\psi_j}$ and $\ket{\psi_R}$\label{table}}
\end{table*}

For each step, we have to apply $H_j$ during a time $\Delta T=\frac{T^\text{ad}}{R^\text{ad}}$, that is the unitary operation:
\beq
U'_j=e^{-iH_j\Delta T}=e^{-i(1-s_j)H_0\Delta T-is_jH_f\Delta T}.
\eeq
As for the analog algorithm, we use the Campbell-Baker-Hausdorff approximation and replace $U'_j$ by
\beq
U''_j=e^{-i(1-s_j)H_0\Delta T}e^{-is_jH_f\Delta T}.
\eeq
The error introduced at each step by this approximation will be
\beq
\opnorm{U'_j-U''_j}\in O(s_j(1-s_j)\frac{\sqrt{N}}{{R^\text{ad}}^2}).
\eeq
For the $R^\text{ad}$ steps we have $U'(T)=\prod_j U'_j$, so that using Eq.~(\ref{opnorm}) gives
\beq
\opnorm{U'(T)-\prod_j U''_j}\in O(\frac{\sqrt{N}}{R^\text{ad}})
\eeq
Consequently, the number of steps $R^\text{ad}=\lfloor\frac{\sqrt{N}}{\epsilon^3}\rfloor$ required in the previous approximation
(i.e., replacing $H(t)$ by $H'(t)$) results in an error of
order $\epsilon^3$ here.

\section{Conclusion}

We have shown that, in spite of their different original implementations, the three quantum search algorithms that have been
found so far are very closely related. They all perform a rotation from the uniform superposition of all states to the solution state at a constant angular velocity,
even though a slightly different path is followed by the analog quantum search algorithm. Their similarities become even more
obvious when they are implemented on a quantum circuit as they all require a number of steps of order $\sqrt{N}$, each step having the same form
$e^{-iH_0\delta t_0}e^{-iH_f\delta t_f}$. Note that the ``duty cycle''
$\delta t_f/(\delta t_0+\delta t_f)$ varies along the evolution according to a specific law in the case of the local adiabatic search algorithm.
Finally, we have shown how one can realize these basic steps on a quantum circuit by using two calls to the
quantum oracle. These results are summarized in Table~\ref{table}.

\begin{acknowledgments}
J.R. acknowledges support from the Belgian foundation FRIA.
N.J.C. is funded in part by the project RESQ under the IST-FET-QJPC European programme.
\end{acknowledgments}

\bibliography{qit}

\end{document}